\documentclass{emulateapj}
\usepackage{apjfonts}
\usepackage{graphicx}
\usepackage{psfig}
\newcommand{\begit}{\begin{itemize}}
\newcommand{\enit}{\end{itemize}}
\newcommand{\begen}{\begin{enumerate}}
\newcommand{\enen}{\end{enumerate}}

\newcommand       \be           {\begin{equation}}
\newcommand       \ee           {\end{equation}}
\newcommand       \bea          {\begin{eqnarray}}
\newcommand       \eea          {\end{eqnarray}}

\newcommand       \kms		{\,{\rm km \,\, s}^{-1}}
\newcommand       \cm		{\,{\rm cm }}
\newcommand       \pc		{\,{\rm pc }}
\newcommand       \yr		{\,{\rm yr }}
\newcommand       \s		{\,{\rm s }}

\newcommand       \g		{\,{\rm g }}
\newcommand       \kpc		{\,{\rm kpc }}
\newcommand       \K		{\,{\rm K }}

\newcommand       \Hz		{\,{\rm Hz }}
\newcommand       \M		{\,{\cal M }}
\newcommand       \eff		{\epsilon_{\rm ff }}
\newcommand       \eg		{\epsilon_{\rm GMC }}

\newcommand       \ergs		{\,{\rm erg \,\, s}^{-1}}
\newcommand       \Rg		{R_{\rm GMC}}
\newcommand       \Mg		{M_{\rm GMC}}

\newcommand       \Mu           {M_{\rm U}}
\newcommand       \Ml           {M_{\rm L}}
\newcommand       \Myr		{\,{\rm Myr }} 

\newcommand       \tff		{\tau_{\rm ff }}
\newcommand       \tms		{\langle\tau_{\rm ms }\rangle}

\newcommand{\beqa}{\begin{eqnarray}} 
\newcommand{\eeqa}{\end{eqnarray}}

\begin{document}

\title{STAR FORMATION EFFICIENCIES AND LIFETIMES OF GIANT MOLECULAR CLOUDS IN THE MILKY WAY}

\author{Norman Murray\altaffilmark{1,2}}

\altaffiltext{1}{Canadian Institute for Theoretical Astrophysics, 60
St.~George Street, University of Toronto, Toronto, ON M5S 3H8, Canada;
murray@cita.utoronto.ca} 
\altaffiltext{2}{Canada Research Chair in Astrophysics}

\begin{abstract}
We use a sample of the 13 most luminous WMAP Galactic free-free sources, responsible for $33\%$ of the free-free emission of the Milky Way, to investigate star formation. The sample contains 40 star forming complexes; we combine this sample with giant molecular cloud (GMC) catalogs in the literature, to identify the host GMCs of 32 of the complexes. We estimate the star formation efficiency $\eg$ and star formation rate per free-fall time $\eff$. We find that $\eg$ ranges from $0.002$ to $0.2$, with an ionizing luminosity-weighted average $\langle\eg\rangle_{Q}=0.08$, compared to the Galactic average $\approx0.005$. Turning to the star formation rate per free-fall time, we find values that range up to $\eff\equiv \tff \cdot \dot M_*/\Mg\approx1$. Weighting by ionizing luminosity, we find an average of $\langle\eff\rangle_{Q}=0.16-0.24$ depending on the estimate of the age of the system. Once again, this is much larger than the Galaxy-wide average value $\eff=0.008$.  We show that the lifetimes of giant molecular clouds at the mean mass found in our sample is $17\pm4\Myr$, about two free-fall times. The GMCs hosting the most luminous clusters are being disrupted by those clusters. Accordingly, we interpret the range in $\eff$  as the result of a time-variable star formation rate; the rate of star formation increases with the age of the host molecular cloud, until the stars disrupt the cloud. These results are inconsistent with the notion that the star formation rate in Milky Way GMCs is determined by the properties of supersonic turbulence.
\end{abstract}


\section{INTRODUCTION}

Giant molecular clouds (GMCs) are stellar nurseries---most stars in most galaxies are born in GMCs. However, the fraction of a GMC's gas that ends up in stars before the cloud is disrupted is rather small, with typical estimates in the Milky Way $\sim2\%$. A related question regards the rate at which stars form in GMCs. Global measurements in both the Milky Way and external galaxies establish that $\sim2\%$ of the molecular gas is turned into star over the galactic dynamical time $R/v_c$, where $R$ is the exponential scale length of the disk and $v_c$ is the circular velocity \citep{1998ApJ...498..541K}. Because gas disks are in hydrostatic equilibrium, the galactic dynamical time is also the disk dynamical time $H/v_T$, where $H$ is the disk scale height and $v_T$ is the turbulent velocity of gas in the disk. Finally, the largest GMCs have diameters that are of order the molecular gas disk scale height, so GMC dynamical times are again of the same order. This implies that only a few percent of the mass of a GMC is turned into stars in the GMC free-fall time.

The lifetimes of GMCs have been a matter of considerable and intense debate, with estimates ranging from a single dynamical or free-fall time \citep{2000ApJ...530..277E,2001ApJ...562..852H}, to a few \citep{2007ApJ...668.1064E}, up to ten or more \citep{1979ApJ...229..578S,2004ASPC..322..245S}, although it is important to note that the last authors were careful to distinguish between the lifetime of a typical $H_2$ molecule (which is likely in excess of ten GMC free-fall times) and that of the GMC of which it is a part.

%\subsection{History}
If GMCs live longer than one or two dynamical times, the question arises as to what supports them. Individual GMCs are observed to have large scale motions of sufficiently high magnitude to support them against their own gravity. However, it is likely that such motions, often referred to as turbulence, decay on a dynamical time. Hence long lived clouds require some energy source if their turbulence is to be maintained. Another possible way to prevent GMC collapse is to appeal to dynamically significant or dominant magnetic fields; such fields (or weaker fields) could be the carriers of MHD turbulence, possibly from sources outside the cloud, which would obviate the need to maintain the turbulence from inside the cloud.

However, it is also possible that GMCs only live for one or two free-fall times, in which case neither magnetic fields nor turbulence are needed to support them; they could be in free fall, as suggested originally by \citet{1974ApJ...189..441G}. These authors pointed out that the similarity between the optically thick CO emission lines and optically thin emission lines from other molecular species in the same clouds could be understood if the clouds had bulk motions with velocities in excess of the thermal linewidths. They argued that there was no clear way to support GMCs, so global collapse was to be expected, leading to the large scale bulk motions needed to explain the observed similarity in line profiles.

This suggestion was disputed by \citet{1974ApJ...192L.149Z}. They give three arguments against free-fall collapse of GMCs. First, they argue that the resulting star formation rate would be too large. They arrive at this conclusion by assuming that upon collapse, all the gas in a GMC would be converted into stars. This ignores any feedback from the stars on the evolution of the cloud. As shown by \citet{2010ApJ...709..424M}, \citet{2010arXiv1004.3290R}, and this paper, there is solid evidence that stellar feedback disrupts the host GMC when a fraction $\eg\approx0.08$ of the GMC mass is converted into stars. This eviscerates the star formation rate argument of \citet{1974ApJ...192L.149Z}, a point also made by \citet{2007ApJ...668.1064E}.

The second argument of Zukerman and Evans is that the line profiles of molecules such as H$_2$CO, which is seen in absorption towards HII regions, should show relative velocity shifts compared to the CO emission lines; the H$_2$CO absorption will be blue shifted if the HII regions are near the centers of the host GMC and the cloud is expanding, since the portion of the cloud between Earth and the HII region is moving towards the observer, while the absorption will be red shifted if the cloud is collapsing. This argument makes the assumption that the absorption lines are formed by gas in the outer regions of the GMC, between the observer and the HII regions, which provide the background continuum against which the H$_2$CO lines are seen. The observations described in \citet{2010arXiv1004.3290R}  establish the presence of outward velocities (in the expanding bubble walls) somewhat in excess of the turbulent velocities measured by CO observations of the GMCs we examine. Despite this, neither double peaked CO emission lines, nor redshifted (relative to the CO lines) absorption lines have been reported towards these star forming complexes. High resolution molecular line observations toward individual bright PAH emission regions might reveal both red and blue shifted CO emission, and associated molecular absorption lines.

%However, the observations also show that the gas surrounding the HII regions is often far from uniform; the free-free emission escapes to distances of order $200\pc$, the radius of the WMAP sources, which \citet{2010ApJ...709..424M} interpret as the ``Extended Low Density'' (ELD) regions identified by \citet{1978A&A....70..565M}. Thus GMCs do appear to ``resemble a piece of Swiss cheese with many holes and little cheese'' \citep{1974ApJ...192L.149Z}. In such a situation, the absorption lines may form a substantial distance from the HII region, and so may not show a substantial infall velocity. 

The third argument of \citet{1974ApJ...192L.149Z} against global infall of GMCs is that there appear to be many infall centers in a given GMC. However, there is no reason to believe that there will be a single infall center in a cloud that is undergoing global collapse: given the clumpy nature of the turbulent ISM, any large clump will act as a local center for collapse of gas in its immediate vicinity.

In this paper we estimate the star formation efficiency, and efficiency per free-fall time, of Milky Way GMCs selected by their free-free luminosity. We show that both efficiencies vary widely, the latter by nearly three decades. Both have large upper limits of order $\sim 0.3$ and $\sim 0.6$ respectively. We also estimate the lifetimes of massive GMCs, finding that they live about one or two free-fall times, consistent with the picture of \citet{1974ApJ...189..441G}.
In \S \ref{sec: efficiency} we give our definition of the two efficiencies $\eg$ and $\eff$. In \S \ref{sec: GMCs}  we describe how we select our star forming complexes, and how we identify their host GMCs.  We estimate the lifetimes of the host GMCs in \S \ref{sec: lifetimes}. We outline the implications of our results and discuss its  relation to previous work in \S \ref{sec: discussion}; in particular we compare our results with the predictions of turbulent star formation theories in section \S \ref{sec: turbulence}. We present our conclusions in the final section.

\section{THE STAR FORMATION EFFICIENCY AND RATE PER FREE FALL TIME OF MILKY WAY GIANT MOLECULAR CLOUDS }
\label{sec: efficiency}
In this section we define and compute two types of star formation efficiency. The first is the efficiency $\eg$ of star formation in GMCs. This is nominally the fraction of a GMC that is converted into stars over the lifetime of a GMC. In fact what we actually measure in this paper is the fraction of a GMC's mass that {\em currently resides} in that GMC in the form of ionizing star clusters. This is because we find stars (and their parent GMCs) by using WMAP to look for free-free radiation, which is powered almost exclusively by massive stars.

\subsection{Star formation efficiency in a GMC}
Star clusters emit ionizing radiation for a rather short time: the 
ionization-weighted (or $Q$-weighted) stellar lifetime is
\be  %$
 \langle t_{\rm ms}\rangle\equiv {1\over \langle q\rangle}
\int_{m_L}^{m_U} q(m_*)t_{\rm ms}(m_*)
{dN\over d\ln m} d\ln m,
\ee  %$
where
\be  \label{eqn: average q}%$
\langle q\rangle\equiv\int_{m_L}^{m_U} q(m_*){dN\over d\ln m} d\ln m.
\ee  %$
The quantity $dN/d\ln m$ is the stellar initial mass function (IMF), which we take to be either that of \citet{2002ApJ...573..366M} modified to have a high mass slope of $-1.35$ rather than their $-1.21$, or the IMF of  \citet{2005ASSL..327...41C}.  With our modification to the Muench et al. slope, the two IMFs are almost identical.  The quantity $\langle q\rangle$ is the number of ionizing photons emitted per second per star, averaged over the IMF. We find $\langle t_{\rm ms}\rangle\approx3.9\Myr$. 

It is worth noting that if a star cluster forms in a time much less than $\langle t_{\rm ms}\rangle$, then that cluster is effectively undetectable via its ionizing radiation after $3.9\Myr$. This is so because the ionizing flux of such a cluster is roughly constant (for slightly less than $\tms$), and then plunges rapidly as the most massive stars in the cluster evolve off the main sequence. In a free-free selected population of such rapidly formed clusters, the average age will be $\langle t_{\rm ms}\rangle/2\approx2\Myr$.

The mass of `live' stars, i.e., the mass in the clusters (of age less than $\tms$) containing the ionizing stars, is given by
%% \footnote{We assume an initial mass function for the stellar population of the form given by \citet{2002ApJ...573..366M} or \citet{2005ASSL..327...41C}; the results using either form are nearly identical.}
%
\be  \label{eq:live mass}%$
M_*= Q \Big / \langle q\rangle / \langle m_*\rangle,
\ee  %$
where
\be \label{eqn: q over m} %$
\langle q\rangle / \langle m_*\rangle = 6.3\times10^{46}
\s\M_\odot^{-1},
\ee  %$
and the IMF averaged stellar mass $\langle m_*\rangle$ is defined in a manner similar to $\langle q\rangle$ in equation (\ref{eqn: average q}).

Since we can measure the mass $M_*$ in live stars, we define the efficiency of star formation as
\be  %$
\eg \equiv {M_*\over \Mg+M_*}.
\ee  %$

As just noted, this stellar mass estimate $M_*$ refers only to stars younger than $\sim3.9\Myr$. If a star cluster takes longer than $\tms$ to form, then the stellar mass associated with that cluster will be larger than that given by equation (\ref{eq:live mass}).  Most GMCs have probably given birth to stars older than $\tms$, so $M_*$ as defined here is a lower limit on the total mass of stars formed in any particular GMC.

\subsection{Star formation efficiency per free-fall time}
The second efficiency we calculate is called the star formation efficiency per free-fall time, $\eff$. For star formation in objects (such as GMCs) of class $X$, this is defined as
\be  %$
\eff\equiv {\dot M_{*X}\over M_X}\tau_{ff-X},
\ee  %$
where $\tau_{ff-X}$ is the free-fall time of objects of class $X$, $M_X$ is the total gas mass in objects of class $X$, and $\dot M_{*X}$ is the star formation rate in such objects \citep{2007ApJ...654..304K}. We restrict our attention in this paper to GMCs. 

We start by computing the Galaxy-averaged value of $\eff$.
The total molecular gas mass in the Milky Way is $M_{\rm tot}\approx10^9M_\odot$, e.g., \citet{1993AIPC..278..267D}. The total ionizing luminosity of the Milky Way is $3.2\times10^{53}\s^{-1}$, eg., \citet{2010ApJ...709..424M} and references therein. 

The star formation rate is given by \citep{1978A&A....66...65S}
\be \label{eq:Mdot} %$
\dot M_*={\langle m\rangle\over\langle q\rangle}{Q\over \langle t_{\rm ms}\rangle}.
\ee  %$
Using equation (\ref{eqn: q over m}) and $Q=3.2\times10^{53}\s^{-1}$, \citet{2010ApJ...709..424M} found $\dot M_*=1.3M_\odot\yr^{-1}$.

As noted above, essentially all stars form in GMCs, and essentially all the molecular gas is in GMCs, so we use $M_X=M_g=10^9M_\odot$. The free fall time for a GMC varies with the GMC mass, but for now we use the value assumed by \citet{2007ApJ...654..304K}, $\tff=4.4\Myr$. The Galaxy wide average star formation efficiency per free fall time is then
\be  \label{eq:eff}%$
\langle\eff\rangle = {\dot M_*\over M_g + M_*}\tff=0.0057.
\ee  %$
This value for $\eff$ is roughly a factor of three smaller than that found by \citet{2007ApJ...654..304K}. They used $\dot M_*=3\,M_\odot\yr^{-1}$, asserting that this was the value found by \citet{1997ApJ...476..144M} (although the latter authors actually found $\dot M_*=4.0\,M_\odot\yr^{-1}$).  \citet{1997ApJ...476..144M} used an IMF with a large mass to light ratio, that of \citet{1986FCPh...11....1S}, which is no longer believed to be a good representation of the actual IMF in the Milky Way.

\section{STAR FORMING GIANT MOLECULAR CLOUDS }\label{sec: GMCs}
In this section we measure the the efficiency, $\eg$, of star formation in individual Milky Way GMCs, and the efficiency of star formation per free fall time, $\eff$. We select the most rapidly star forming GMCs in the Galaxy: the thirty two GMCs we select are responsible for $31\%$ of the star formation in the Galaxy.

We start with the WMAP free-free thermal radio fluxes of sources found by \citet{2010ApJ...709..424M}. Examination of Spitzer and MSX images in the direction of the WMAP sources reveals one to several `star forming complexes' in the direction of each of the WMAP sources \citep{2010arXiv1004.3290R}. A star forming region is identified on the basis of the morphology in the $8\mu$m Spitzer and MSX images, combined with measurements of radio recombination line (or molecular line) radial velocities taken from the literature. The result is a list of 40 star forming complexes with Galactic coordinates and radial velocities, $(l,b,v_r)$. These complexes have characteristic sizes of $\sim 25\pc$, ranging between $2\pc$ and $70\pc$ (see table 1).

Once all the star forming complexes are identified, we divide the free-free flux in each WMAP source between the star forming complexes contained in that source. In two cases, G283 and G327, there is only a single star forming region in the WMAP source. We assign the entire free-free flux of the WMAP source to that star forming region; in both cases the size of the star forming region (determined by the extent of the region outlined by the radio recombination line velocity measurements) is similar to that of the WMAP source, supporting the view that only that source is responsible for the free-free emission seen by WMAP in that direction. 

In four of the WMAP sources there are exactly two star forming complexes; we split the flux between the star forming complexes based on the radio recombination line fluxes associated with each source. Such a division is a rather crude approximation; ground based radio continuum fluxes toward each of these complexes are systematically lower than the continuum fluxes measured by WMAP, but the flux ratios vary unpredictably from object to object. In its favor, this recombination line ratio split was consistent with the relative amount of $8\mu$m flux associated with the star forming complexes (accounting for the fact that the total $8\mu$m flux appears to scale as the square of the free-free flux, \citet{2010arXiv1004.3290R}).
 
\citet{2010arXiv1004.3290R} identified four star forming complexes in G10, three in G34 and G311, and five in G337. In all these cases the star forming complexes are well separated in $(l,b,v_r)$ space, as was the $8\mu$m emission in $(l,b)$. The division of the free-free emission amongst the sources was made in the same manner as for the previous cases.

The two WMAP sources G24 and G30 are very confused, with eight and six star forming complexes, respectively. We again used the recombination line ratios to assign the free-free flux to each source, but we have less faith in the results; to indicate this, we plot these points using open polygons, as opposed to the filled polygons plotted for the less confused WMAP sources.

Having assigned fluxes to each star forming region, we then use the kinematic distance $D$ to that region to calculate the free-free luminosity $L_\nu=4\pi D^2 f_\nu$ emitted by that region, and the rate $Q=1.33\times10^{26}L_\nu\s^{-1}$ \citep{2010ApJ...709..424M} of ionizing photons emitted by each source per second required to power the observed free-free luminosity. In cases where the distances listed in \citet{2010arXiv1004.3290R} and the relevant GMC catalog do not agree, we use the distance of the former. In the case of Rahman \& Murray sources 5, 7, 9, 12, 15, and 35, where those authors do not resolve the distance ambiguity, we use the distance that is closer to that of the relevant GMC catalog distance.

Following \citet{1997ApJ...476..144M} we then increase the rate of ionizing photons by a factor $1.37$ to account for the fact that some of those photons are absorbed by dust, and hence do not contribute to the free-free emission detected by WMAP.

The live stellar mass was then calculated using eqn. (\ref{eq:live mass}); a given GMC almost certainly has given birth to stars older than $\langle t_{\rm ms}\rangle$, so $M_*$ as defined here is a lower limit on the total mass of stars formed in the GMCs we examine. 

The next step in the calculation of $\eg$ and $\eff$ is to identify the host GMCs of our star forming complexes. We search the GMC catalogs of \citet{1987ApJ...319..730S,1988ApJ...331..181G,1989LNP...331..139B}, and \citet{2009ApJ...699.1092H} for objects having centers at the same Galactic longitude within
$\pm0.4$ degrees, Galactic latitude within $\pm0.4$ degrees ($\sim50\pc$ at $D=10\kpc$, comparable to the radius of a typical cloud), and the same heliocentric radial velocity within $\pm20\kms$.

We find matches for 32 out of our 40 star forming complexes (SFCs); results are given in table 2. A number of GMCs are listed in both 
\citet{1987ApJ...319..730S} and \citet{2009ApJ...699.1092H}, in which case we use the values in the latter. We corrected the GMC radii to account for the difference between our assumed value for the distance to the Galactic center $R_0=8.5\kpc$ and that used in the GMC surveys. We have followed  \citet{1997ApJ...476..166W} (their table 1) in correcting the masses of the GMCs in the first three surveys. Heyer et al. use the same distance to the Galactic center we do ($8.5\kpc$); we use $\Mg=2\times M_{\rm LTE}$, as those authors recommend (corresponding to $X_{CO}=1.9\times10^{20}\cm^{-2} (K \kms)^{-1}$).

The average star-forming GMC mass in our ionizing luminosity-selected sample is $1.5\times10^6M_\odot$. The ionizing luminosity-weighted GMC mass is $2.3\times10^6M_\odot$. The average and Q-weighted GMC radii are $58\pc$ and $76\pc$ respectively. Similarly, the average and Q-weighted free-fall times are $8\Myr$ and $9\Myr$.

These averages include the confused GMCs associated with SFCs 5 to 18. These GMCs have been investigated by both \citet{1987ApJ...319..730S} and \citet{2009ApJ...699.1092H}; the masses of these GMCs differ between the two studies (as noted above, we use the results of \citet{2009ApJ...699.1092H} where possible). While the average properties of the confused sources (and their host GMCs) are similar to those of the un-confused sources, the measured properties of individual confused sources may be unreliable. For example, source 10 has $M_*=2.3\times10^4M_\odot$ and $\Mg=1.6\times10^4M_\odot$. As indicated in table 2, the force of radiation exceeds that of self-gravity by more than one hundred; since the bubble is not expanding particularly rapidly, this extreme force ratio seems unlikely. We treat this star forming region as an outlier, since we believe that is probably miss-identified, and do not include it our statistics.

If we confine our attention to un-confused GMCs, the average and $Q$-weighted radii are $91\pc$ and $116\pc$. Similarly, the average and $Q$ weighted free-fall times for the un-confused GMCs are $11\Myr$ and $13\Myr$.

\begin{figure}
%\epsscale{0.9} 
%\plotone{Fig_lockup.eps}
%\plotone{f2.eps}
\includegraphics[width=\hsize]{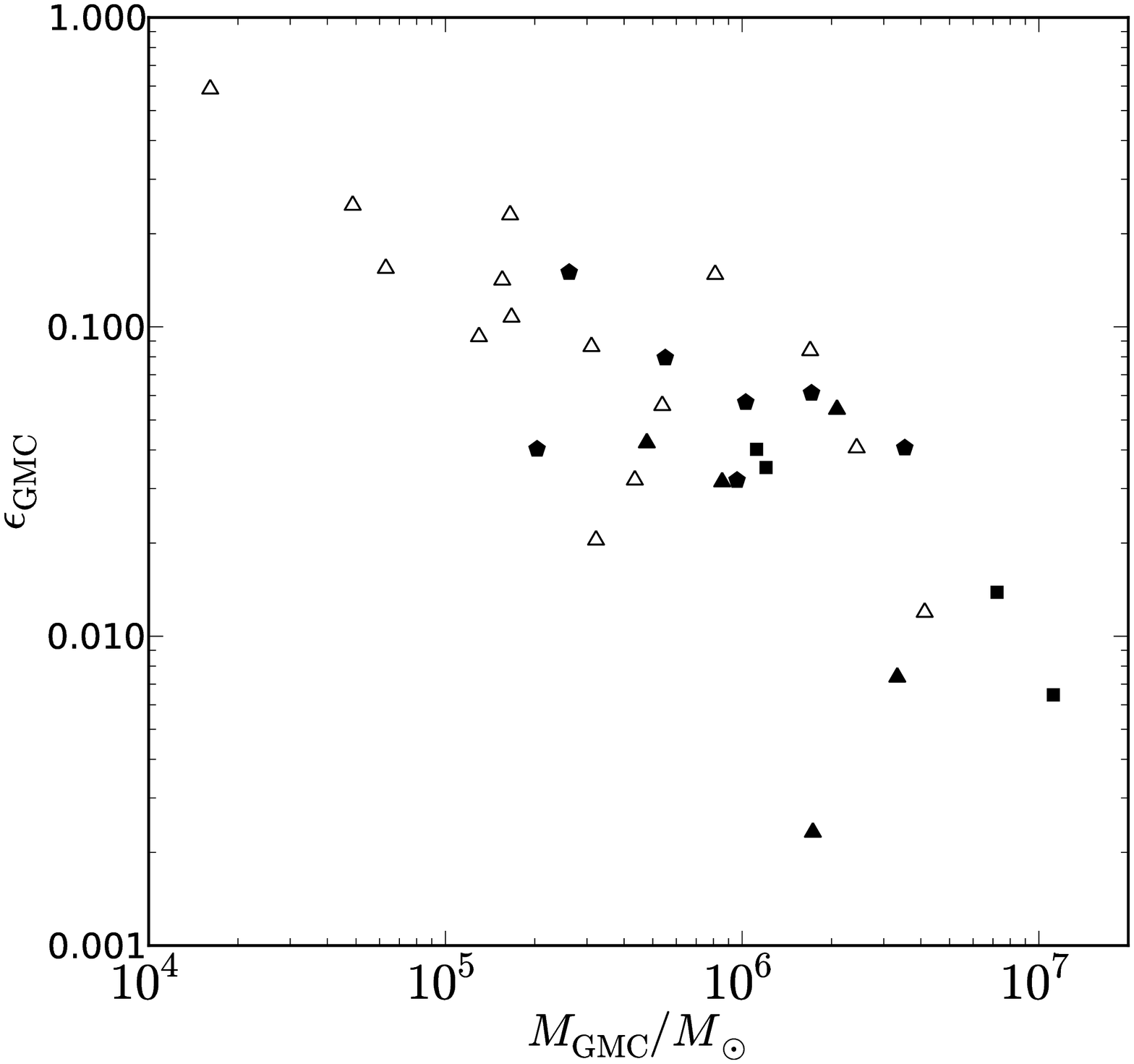}
\caption{%
The ratio $\eg\equiv M_*/(\Mg+M_*)$ of the mass of young ($t<\tms$) stars in star-forming Milky Way GMCs of mass $\Mg$. Filled triangles are for GMCs in \citet{1987ApJ...319..730S}, open triangles (most of which correspond to objects in the confused SFCs G24 and G30) are for GMCs in \citet{2009ApJ...699.1092H}, filled squares are for GMCs in \citet{1988ApJ...331..181G}, and filled pentagons are for GMCs
 in \citet{1989LNP...331..139B}. The stellar masses are found from WMAP free-free fluxes as described in the text. We find $\langle\eg\rangle=0.09$, while the $Q$-weighted average is $\langle\eg\rangle_{Q}=0.08$.
}
\label{fig:epsilonG}
\end{figure}

Having found the mass of the host GMCs, we can estimate the star formation efficiency. The results are plotted in Figure~\ref{fig:epsilonG}. 

The sample average $\langle\eg\rangle=0.09$ for our 32 star forming complexes.  We stress again that $\eg$ as defined here is a lower limit to the total star formation efficiency of the host GMC. Any stars in clusters older than $\tms$ will not be detected by WMAP and hence are not included in our estimate of $M_*$.

This average is a bit higher than that usually quoted (typical estimates are more like $0.02$), but we have selected the most active star forming complexes in the Galaxy, rather than a Galaxy wide average.

The Q-weighted efficiency is defined as $\langle \eg\rangle_{Q}\equiv\Sigma_i Q_i \epsilon_{G,i}/(\Sigma_i Q_i)$; we find $\langle \eg\rangle_{Q}=0.08$.

\subsection{The star formation rate per free-fall time}
Since we know both the mass and radius of the host GMC, we can calculate the GMC free fall time $\tff\equiv\sqrt{3\pi /(32 G\bar\rho)}$, where $\bar \rho \equiv 3\Mg/(4\pi \Rg^3)$.  We can then find the star formation efficiency per free fall time. However, we cannot use equation (\ref{eq:eff}) directly, since the star formation in GMCs is not steady state: it is possible, or even likely, that most of the star formation takes place on a time scale shorter than $\langle t_{\rm ms}\rangle$. 

Instead, we combine equations (\ref{eq:Mdot}) and (\ref{eq:eff}):
\be  %$
\eff={M_*\over \Mg + M_*}{\tff\over\langle t_{\rm ms}\rangle}.
\ee  %$
Note that $\tff=1.65\times10^4R_{pc}^{3/2}/M_6^{1/2}\yr$.

\begin{figure}
%\plotone{f2.eps}
\includegraphics[width=\hsize]{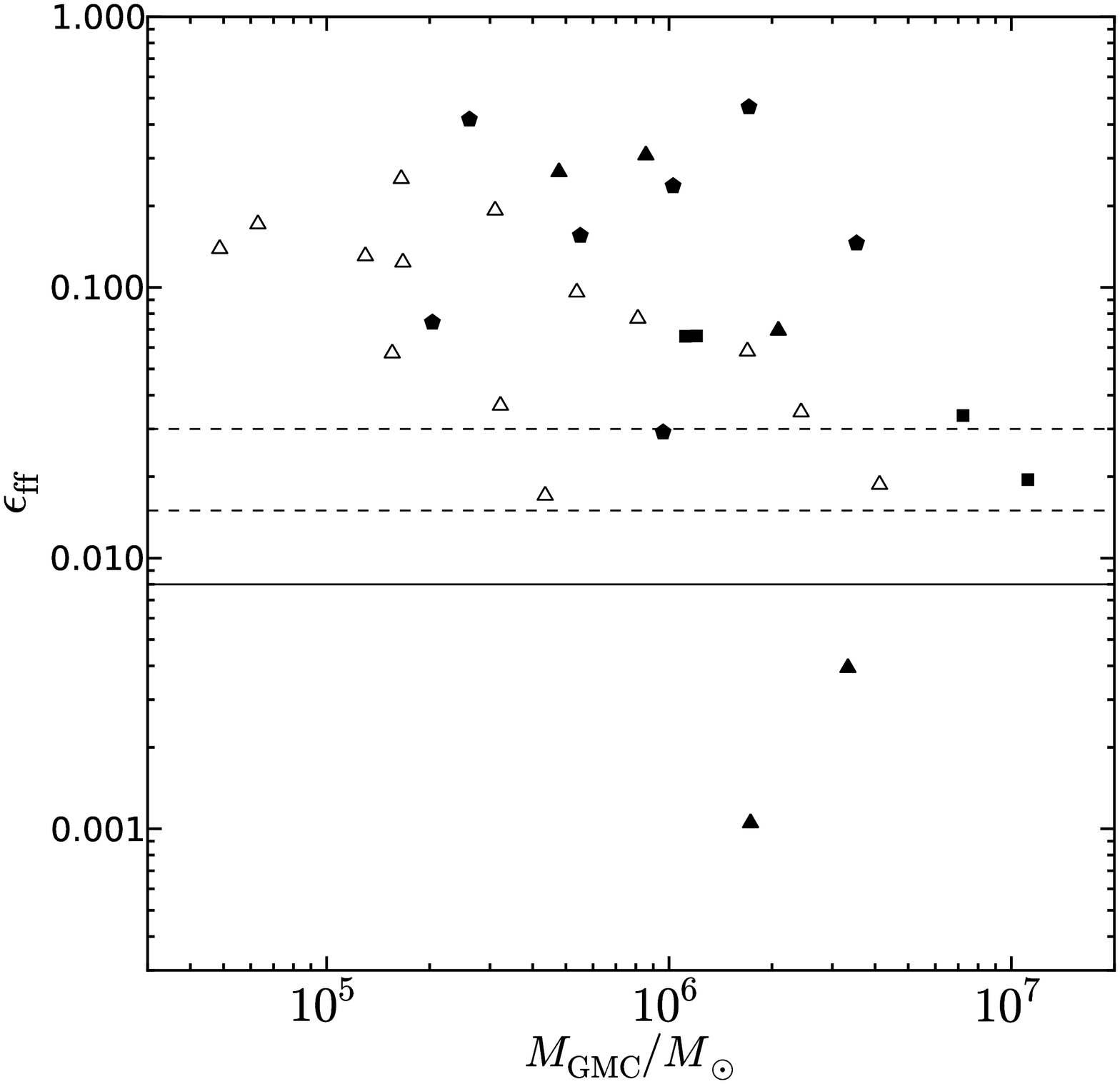}
\caption{%Fig. 2---
The star formation rate per free fall time,  $\eff\equiv \left[ M_*/(\Mg+M_*)\right] (\tff/\tms) = \eg (\tff/\tms)$ of free-free selected GMCs of mass $\Mg$.  Symbols as in Figure 1. The sample average $\langle\eff\rangle=0.16$ while the $Q$-weighted $\langle\eff\rangle_Q=0.15$. The solid horizontal line shows the Milky Way average value of $\eff=0.008$, while the two dashed lines show the range around $\eff\approx 0.02$ discussed by \citet{2005ApJ...630..250K} and \citet{2007ApJ...654..304K}. Most of the star formation in the Milky Way occurs in very rapid bursts, in which $\sim8\%$ of the host GMC is converted to stars in half (or less) of the free-fall time of the GMC.
\label{fig:epsilonff}}
\end{figure}

%% The ionizing flux of a single age drops precipitously at an age of $\tau_Q=3.9\Myr$, so one estimate of $\eff$ can be obtained by using $\dot M_*=M_*/\tau_Q$, or
%% %
%% \be  %$
%% \eff={M_*\over \Mg}{\tff\over \tau_Q}.
%% \ee  %$
%% % 
The results using this estimate are given in Figure \ref{fig:epsilonff}, which shows that $\eff$ ranges from $\approx10^{-3}$ to $1.2$ (the latter indicating that the entire host GMC mass would be converted into stars, in the absence of feedback effects, in slightly less than a free fall time), with a mean $\langle\eff\rangle=0.16$.

The $Q$-weighted efficiency per free-fall time is
\be  %$
\langle\eff\rangle_Q=0.15
\ee  %$
similar to the sample average value.

Once again, these are lower limits for $\eff$, since not all the ionizing clusters are $3.9\Myr$ old. An arguably more realistic but still simple estimate is to take a mean cluster age of $\tau_Q/2$ (this assumes that the clusters form in a time short compared to $3.9\Myr$) leading to an estimate of $\langle\eff\rangle=0.3$.

A third estimate uses the smaller of the dynamical time $\tau_{\rm dyn}=R/v$ and $\tau_Q$. We argue that our dynamical times are likely to be overestimates, since the radial velocity spread is a lower limit to the true bubble expansion velocity. We use the smaller of $\tms$ and the dynamical time
\be  %$
\eff={M_*\over \Mg + M_*}{\tff\over {\rm min}(\tms,\tau_{\rm dyn})}.
\ee  %$
%
%The results are shown as open squares in Figure (\ref{fig:epsilonff}). 
Using this estimate we find
\be  %$
\langle\eff\rangle_Q=0.24,
\ee  %$
for a $Q$-weighted average. 

\section{GMC LIFETIMES} \label{sec: lifetimes}
The result that $\sim30\%$ of the star formation in the Milky Way occurs in 32 GMCs is truly remarkable. This becomes apparent when we compare  the total molecular gas mass in the Milky Way, $M_{\rm tot}\approx10^9M_\odot$  \citep{1993AIPC..278..267D}, to the mass in the $32$ star forming GMCs, with total gas mass $M\approx 5.9\times10^7M_\odot$; $30\%$ of the star formation takes place in clouds that contain only $6\%$ of the molecular gas mass. 

All the star forming complexes we examine contain expanding bubbles, with typical expansion velocities of $\sim 10\kms$ and radii $R_b$ ranging from $3\pc$ to $100\pc$ \citep{2010arXiv1004.3290R}. In the more  vigorously star forming GMCs, the outward force exerted by radiation from the stars exceeds the inward force of gravity acting on the GMC; we use
\be %$
F_{\rm rad}=L/c=\xi Q/c,
\ee %$
for the force due to radiation, where $\xi=8\times10^{-11}\ergs \cdot\s$, appropriate for our choice of IMF. The outward force due to the pressure of the ionized gas is estimated as $F_{\rm gas}=4\pi R_b^2P$, where $P$ is the gas pressure. We use the expression $P= nkT$, where $k$ is Boltzmann's constant, the temperature of the ionized gas is $T=7000\K$, and 
\be  %$
n=\sqrt{3Q\over 4\pi R_b^3 \alpha_{rec}},
\ee  %$
where $\alpha_{rec}$ is the recombination coefficient. 

\begin{figure}
%\plotone{f3.eps}
\includegraphics[width=\hsize]{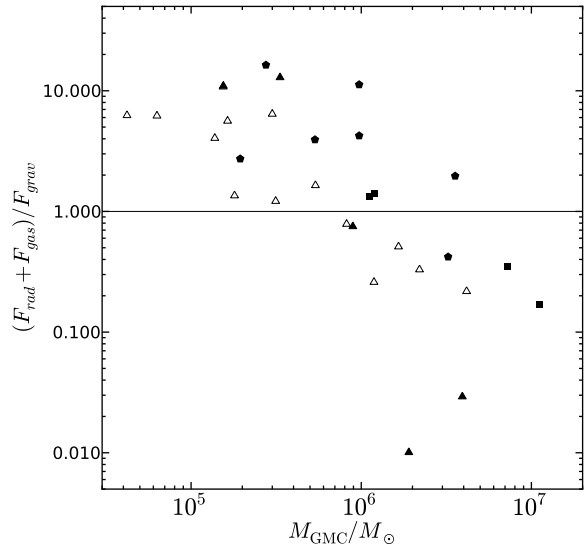}
\caption{%%
The ratio of outward (radiation and gas pressure) forces to the self-gravity force of the GMCs, $(F_{\rm rad}+ F_{\rm gas})/F_{\rm grav}$, as a function of the host GMC mass. Twelve of the eighteen unconfused clusters (shown as filled polygons) have ratios larger than unity, indicating that the star clusters should be blowing expanding bubbles, consistent with the 8 micron bubble morphology seen in Spitzer or MSX images of the star forming complexes. In almost all cases the radiation pressure is larger than the gas pressure, with the average  ratio $F_{\rm rad}/F_{\rm gas}\sim2$. 
\label{fig:force}}
\end{figure}

For the force of gravity we estimate
\be %$
F_{\rm grav}= {G\Mg\Mg\over \Rg^2}.
\ee %$
The ratio $F_{\rm rad}/F_{\rm grav}$ is given in column 9 of table 2, and the ratio $(F_{\rm rad}+F_{\rm gas})/F_{\rm grav}$ is plotted in Figure \ref{fig:force}. More than half of the unconfused sources have a total outward force larger than the force of self-gravity. This is consistent with our interpretation of the radial velocity spreads as being due to expansion of the bubble walls, and strongly suggests that the star clusters are disrupting their host GMCs. We note that in almost all our sources, $F_{\rm rad}>F_{\rm gas}$, as expected for such massive clusters \citep{2010ApJ...709..191M}.
%

%% We have argued previously that the star formation efficiency $\eg$ should increase with increasing surface density $\Sg$ of the host GMCs \citet{2010ApJ...709..191M}. Figure \ref{fig:Sigma} plots $\eg$ against $\Sg$. There is no obvious trend of increasing $\eg$ with increasing $\Sg$

%% %
%% \begin{figure}
%% \includegraphics[width=\hsize]{f4.eps}
%% \caption{%%
%% The star formation efficiency $\eg$ versus the surface density $\Sg$ of the host GMC. The solid line is the lower limit predicted by \citet{2010ApJ...709..191M}.
%% \label{fig:Sigma}}
%% \end{figure}
%% %

The largest bubbles appear likely to disrupt the host GMC; the bubble walls have enough momentum to sweep up the rest of the gas in the host cloud, driving it to $r\sim 200\pc$, at which point tidal shear will complete the disruption. We see these clouds in their death throes. 

We can estimate the lifetimes of massive ($\Mg\gtrsim10^6M_\odot$) Milky Way GMCs against disruption by the effects of the star clusters that form in them. To do so we assume that the massive GMCs harboring the massive clusters we examine here are drawn from the massive GMC population as a whole. This assumption implies that all massive GMCs will eventually form massive star clusters, but leaves open the possibility that less massive GMCs (say with masses below $\sim10^5M_\odot$) do not necessarily form many stars. For example, cloud fragments from the objects we have found, which may well have masses as large as $10^5M_\odot$, may not be self-gravitating. If they are not, they may avoid any substantial star formation until the next time they find themselves inside a (larger) gravitationally bound object.

To proceed with our estimate of massive GMC lifetimes, we need to estimate the number of GMCs in the parent population of our star forming objects. We do so in two different ways. First, we estimate the number of GMCs in the Milky Way that are required to produce a fraction $f_*\approx0.31$ of the observed total star formation rate, $dM_*/dt=1.3\,M_\odot\yr^{-1}$. Second, we estimate the number of clouds with masses above some characteristic mass, either the average or $Q$-weighted mean GMC mass.

\begin{figure}
\plotone{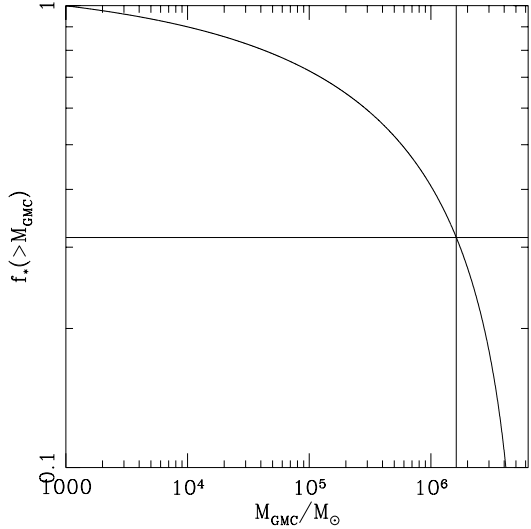}
\caption{%%
The cumulative fraction $f_*(>M)$ of the Galactic star formation rate produced in GMCs with $\Mg>M$. The fraction $f_*=0.31$ is produced by GMCs with masses larger than $\Mg=1.6\times10^6M_\odot$.
\label{fig:fstar}}
\end{figure}

To estimate the star formation rate in clouds with masses greater than some mass $\Mg$, we use the Galaxy-wide average star formation rate per free-fall time, and integrate from the most massive clouds (mass $\Mg$) down towards the least massive clouds, with mass $\Ml$. The number of clouds follows a power law with index $\alpha\approx1.5-1.6$ \citep{1984A&A...133...99C,1986ApJ...305..892D,
1996ApJ...458..561D,2009ApJS..182..131R}. Let $f_*$ be the fraction of the Galactic star formation rate that occurs in clouds with mass $M_{f*}$ or larger. We show in the appendix that 
\be  %$
M_{f_*}=\left[1-f_*\left(1-\left({\Ml\over\Mu}\right)^\delta\right)\right]^{1/\delta} \Mu,
\ee  %$n
where $\delta = (7/4)-\alpha$. The upper and lower limits on the GMC masses are denoted by $\Mu=6\times10^6M_\odot$ and $\Ml=10^3M_\odot$, respectively; varying $\Ml$ has little effect on the result. Figure \ref{fig:fstar} shows the cumulative fraction of the star formation rate produced in GMCs with masses greater then $M$.

\begin{figure}
\plotone{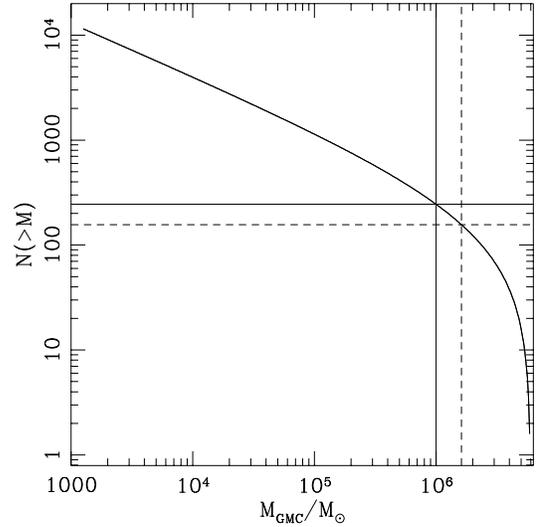}
\caption{%%
The cumulative number of GMCs in the Milky Way, plotted as a function of GMC mass $\Mg$. The total number of clouds is $N\approx 13,000$. The vertical dashed line is at $\Mg=1.6\times10^6M_\odot$, corresponding to the mass of GMCs that produce $31\%$ of the star formation in the Milky Way; there are $\sim 160$ such clouds. The vertical solid line at $\Mg=1.5\times10^6M_\odot$ corresponds to the average GMC mass in our sample; there are $\sim 170$ GMCs in the Milky Way having masses this larger or larger. These clouds contain $\sim50\%$ of the molecular gas in the Galaxy.
\label{fig:dndm}}
\end{figure}

Figure \ref{fig:dndm} shows the cumulative number $N(>M)$ of GMCs with masses $\Mg>M$. The number of Milky Way GMCs with masses larger than $M(f_*=0.31)=1.6\times10^6M_\odot$ is $158$, so for every star forming GMC in our sample, there are an average of $4.9$ GMCs in the parent population. We use the relation 
\be  %$
\tau_{GMC} = {N_{GMC}(>M_{f_*})\over N_{GMC,Q}(f_*)}\tms,
\ee  %$
where we assume that the average lifetime of the star clusters in our sample is $\tms=3.9\Myr$. We find that the lifetime of the parent GMCs is $\sim19\Myr$. The $Q$ weighted average free-fall time of these GMCs is $\approx9\Myr$, and $10\Myr$ for un-confused GMCs, so our first estimate is that massive GMCs live  $1.7-2.1$ free-fall times before they are disrupted by expanding bubbles produced by the star clusters they contain.

We make a second estimate for the lifetimes of massive GMCs by finding a characteristic GMC mass for the parent GMCs in our ionizing luminosity-selected sample of star forming complexes. We then compare the number of GMCs in our sample to the total number of GMCs at or above that characteristic mass.

The mean and $Q$-weighted mean GMC masses of our sample are $1.5\times10^6M_\odot$ and $2.3\times10^6M_\odot$. The corresponding $N(>1.5\times10^6M_\odot)=169$ and $N(>2.3\times10^6M_\odot)=104$. The estimated GMC lifetimes range between $13-21\Myr$, or $1.4-2.3\tff$, similar to our first estimate. We take the lifetime of massive GMCs in the Milky Way to be $1.9\pm 0.4\tff$.

The ionizing luminosity of some of our star forming complexes is rather low, while our original WMAP sample was chosen for its high luminosity. The confusion of the WMAP sources results in the inclusion of low luminosity SFCs, which in turn leads to our over-counting the number of GMCs in a true luminosity-selected sample. We can try to correct for this by making a cut in $Q$, at a value we will denote $Q_{\rm cut}$. When we do so, we find that the average GMC mass increases as a function of $Q_{\rm cut}$. So while it is true that increasing $Q_{\rm cut}$ decreases the number of star-forming GMCs in our sample, which tends to increase the estimated lifetimes of the GMCs, at the same time $\langle \Mg\rangle(Q_{\rm cut})$ also increases, which tends to decrease the estimated lifetimes of the GMCs. The net result is that the ratio $N_{\rm GMC}/N_{\rm cut}$ increases slowly, from $4.6$ to $6.1$, and the estimated lifetime increases by $30\%$ by the time $Q_{\rm cut}=5\times10^{51}\s^{-1}$; at this value of $Q_{\rm cut}$, there are only $7$ sources left in the sample, and the mean mass is no longer well defined, i.e., the mean GMC mass begins to drop with increasing $Q_{\rm cut}$ beyond this value.

\section{DISCUSSION}\label{sec: discussion}

Figure \ref{fig:force} shows that the sum of radiation pressure and gas pressure forces exceeds the self gravity in 18 of our 31 GMCs (excluding the probable misidentification), indicating that these GMCs should be in the process of being disrupted. The presence of expanding bubbles in these systems is consistent with this notion, and with the predictions of \citet{2010ApJ...709..191M}. This strongly suggests that these GMCs are at the end of their lives, either as converging flows or as gravitationally bound objects.

Both the observations and the theory of \citet{2010ApJ...709..191M} suggest that the process of disruption takes about one half of a free fall time, since the expansion velocity of the bubbles is $v\sim10-15\kms$, about a factor of two larger than $v_{\rm GMC}$. The time for turbulence to decay (assuming the clouds are initially turbulence supported) is about one free fall time, so the time for an unsupported GMC to collapse is $\sim 1-2\tff$. The total life time of an unsupported GMC, from formation through collapse and star formation to disruption, would then be $1.5-2.5\tff$, similar to the value we measure. In other words, the fact that massive GMCs in the Milky Way appear to live only about two free-fall times is consistent with the suggestion of \citet{1974ApJ...189..441G} that all GMCs are in the process of collapse; the short lifetimes of massive GMCs obviate the need for any means of support, including driven turbulence or magnetic support. 

As noted in the introduction, the argument of \citet{1974ApJ...192L.149Z}, that rapid collapse of GMCs would lead to a star formation rate in the Milky Way that was much higher than the observed rate, does not apply if the stars that are formed disrupt the host GMC. We have shown when $\eg\sim0.08$ the GMCs we examine appear to be in the process of disruption. This suggests that the star formation rate is much less than $\Mg/\tff$ because only a small fraction of $\Mg$ is converted into stars before the cloud is disrupted by radiation pressure. 

While our results are consistent with free-fall collapse of GMCs, they do not require such a collapse. It may be, for example, that the bulk of the material in GMCs is held up by magnetic fields, but that over one dynamical time ($\sim 10\Myr$), some $10-15\%$ of the GMC accretes onto one or two massive clumps. Roughly half of this material will form stars, with the rest dispersed back in to the ISM, resulting in the observed $\langle\eg\rangle_Q\sim0.08$. It may also be that the GMCs are not gravitationally bound to begin with; measurements of the virial parameters of massive clouds are near unity.

The last statement is consistent with the short lifetimes we have inferred for massive GMCs---dissipating sufficient energy to make a strongly bound GMC would require more than a single free-fall time. One could also imagine that some supply of energy could maintain a cloud with a virial parameter near unity. However, since the clouds only live for one or two free-fall times, it is difficult to envision a steady state arising, in which decay of turbulent energy is balanced by some source of energy. 

Adding up the star formation from all the GMCs, $\sum\eg\Mg/(2\tff)$ (the factor two comes from our estimate of the GMC lifetime) the star formation rate is only a factor of $\sim2$ larger than the observed Milky Way star formation rate. This implies that the fragments of the disrupted GMCs must reassemble in $\sim 2\tff$; if they did not, the star formation rate would be lower than observed. In addition, if the fragments of the disrupted $10^6M_\odot$ GMCs do not reassemble after a short time, more of the molecular gas would reside in smaller clouds, contrary to observations. Both these observations strongly suggest that the cloud fragments reassemble in less than an orbital period.

We note that the lifetimes of the majority of massive GMCs cannot be much larger than a few $\tau_{\rm dyn}$, for two reasons: first, we would see massive GMCs between spiral arms, something for which there is little observational evidence. Second, the star formation rate would be much less than that observed. In other words, the majority of the most massive GMCs in the Milky Way cannot be supported against collapse by any mechanism that suppresses star formation for more than two or three GMC free fall times.

\subsection{Dynamical star formation}

\citet{2001ApJ...562..852H} and \citet{2007RMxAA..43..123B} argue that GMCs and stars form rapidly, in a dynamical time or less, based on observations of star forming regions within $1\kpc$ of the Sun. The observational argument is simple: most of the GMCs in the solar vicinity have substantial star formation, so that the age of the star clusters and that of the GMCs must be similar. Since the average age of the star clusters is of order $1-3\Myr$, they conclude that the typical GMC lifetime is also $\sim1-3\Myr$.

The different (short) lifetimes for the host GMCs found by these authors, compared to the lifetimes found here, arise from several causes. First is the observational fact that they find a ratio of total to star-containing GMCs of 1.5, compared to our ratio of free-free dim to free-free luminous objects of $\sim4$, for a total to luminous ratio of $\sim5$. By itself, this difference accounts for a ratio of more than three in the estimated GMC lifetime.

It may well be the case that most of the Milky Way clouds with $\Mg>1.5\times10^6M_\odot$ (the parent population of our sample) host stars, even ionizing stars, that have $Q$ below our selection criteria. If we counted such clouds then we might find a ratio similar to that found by \citet{2007RMxAA..43..123B}. However, it would not be correct to conclude that the GMC lifetime was then $1.5$ times $\tms$, unless all the GMCs were in the process of being disrupted (by some mechanism not related to radiation pressure or HII gas pressure, since either would not be adequate in such dim sources).

A second source for the difference in estimated GMC lifetime arises because \citet{2007RMxAA..43..123B} employ an {\em average age} for their stars of $\sim2\Myr$, with an upper limit of $5\Myr$, while we use a {\em total lifetime} of $\tms=3.9\Myr$. We believe that the correct procedure is to use the total (embedded) lifetime of the stellar tracers, rather than the average age. We add the qualifier {\em embedded} since the host GMC may be disrupted before the life of the stellar tracers employed is reached. The total embedded lifetime is in general longer than the average embedded lifetime, so using the latter will underestimate the GMC lifetime. Using their upper limit of $5\Myr$ (given in their section 3) as the total embedded lifetime would result in an increase in their estimated GMC lifetime by a factor of $2.5$, to about $5\Myr$. If we then multiply by the ratio of total to star-containing GMCs ($1.5$), the GMC lifetime in their sample is $\sim 7.5\Myr$, somewhat larger than a third of the $\sim19\Myr$ lifetime found in our sample. 

There is also a physical effect that may contribute to the difference in estimated GMC lifetimes found by \citet{2007RMxAA..43..123B} and those found here. Table 1 of \citet{2007RMxAA..43..123B} lists 21 small GMCs within $1\kpc$ of the Sun, containing a total of $4.2\times10^6M_\odot$ of gas, of which 14 show some signs of star formation. 
%% The bulk of the gas mass is in three clouds each having $\Mg\approx 8\times10^5M_\odot$, two of these being the Cyg Rift and the Cyg OB7 cloud. The total mass of live stars in the latter is small compared to most of the objects we examine in this paper (The entire Cygnus region has $Q\approx2.7\times10^{51}\s^{-1}$, Orion has only $Q=10^{49}\s^{-1}$). 
The average GMC mass in their sample is $2\times10^5M_\odot$.

The GMCs considered here are much more massive (average $\Mg=15\times10^5$ versus $2\times10^5M_\odot$) in our sample, much larger, and more diffuse, so the GMCs in our sample have longer free-fall times than the GMCs in \citet{2007RMxAA..43..123B}. The more massive clouds will naturally take longer to form and collapse than less massive clouds. If the less massive clouds are destroyed by the same mechanism (a combination of radiation and gas pressure), then the larger clouds will live a factor $M^{1/4}\sim 1.6$ longer, or $\sim12\Myr$,  (although they may survive for the same number of free-fall times).

We conclude that the GMC lifetimes, measured in terms of free-fall times, found by \citet{2007RMxAA..43..123B} are consistent with those we find, given the different GMC masses considered and the uncertainties involved.

\subsection{Turbulence and star formation}\label{sec: turbulence}
The high fraction of gas turned into stars in a free fall time in star forming Milky Way GMCs ($\sim 16-25 \%$) is surprising in light of recent theories invoking turbulence to regulate the rate of star formation, e.g., \citet{1995MNRAS.277..377P,2005ApJ...630..250K}. In these theories, the rate of star formation is related to the amount of gas above a critical density, set by the requirement that the local Jeans length be less than the sonic length (the length at which the velocity of turbulent motions equals the sound speed of the gas). The critical density depends on the Mach number ${\cal M}\equiv \sigma/c_s$ and the virial parameter
\be  %$
\alpha_{\rm vir}\equiv {5\sigma^2 \Rg\over G\Mg},
\ee  %$
where $\sigma$ is the 1-D line of sight velocity dispersion. 

We have calculated both ${\cal M}$ and $\alpha_{\rm vir}$ for the host GMCs in our sample, and in the parent GMC population, finding values clustering about ${\cal M}\approx 10-15$ and $\alpha_{\rm vir}\approx 0.3-2$, with the lower values for $\alpha_{\rm vir}$ seen in the more massive GMCs, those with  $\Mg>10^6M_\odot$.\footnote{We note that the sample in \citet{1989LNP...331..139B} has anomalously high values of $\alpha_{\rm vir}$, apparently driven by the high linewidths reported there. Despite this, the values of $\eg$ and $\eff$ are similar to those found for the other populations we consider.} The average virial parameter for the Heyer et al. sample objects with $\Mg>10^6M_\odot$ is $\alpha=0.6\pm0.3$; the subsample with high free-free luminosities has an average $\alpha=0.9$. While not highly significant, this is consistent with the notion that the host GMCs in the latter objects are being disrupted by their daughter star clusters. 

In calculating ${\cal M}$ we adopted $T=15\K$, and a mean molecular weight of $3.9\times10^{-24}\g$. We then used these values in eqn. (30) of \citet{2005ApJ...630..250K} to find the predicted value of $\eff\approx0.02$. This is roughly a factor ten lower than the average $\eff$ in our luminosity selected sample.

\citet{2007ApJ...654..304K} have estimated $\eff$ in different environments in the Milky Way, finding values ranging from $0.013$ for GMCs and other diffuse objects, to $0.2$ (their CS(5-4) point). As noted above, we find a lower value for the Milky Way average GMC (due to our lower estimated star formation rate) but a much higher value for actively star forming GMCs, which we infer to be at the end of their lives. One natural interpretation is that the star formation rate per free fall time varies with time in a given GMC, increasing rapidly just before the cloud is disrupted. Since there is no indication that the Mach number or virial parameters of our rapidly star forming GMCs differ dramatically from the sample mean values (and the Mach number dependence of $\eff$ in the turbulence picture depends only weakly on ${\cal M}$ in any case), this suggests that turbulence is not, by itself, controlling the rate of star formation in Milky Way GMCs.

This point is worth stressing. \citet{2005ApJ...630..250K} estimate the upper and lower limits to the turbulence limited star formation rate, finding values a factor $10$ below their best estimate to $6$ times above. They state that a more realistic estimate is a factor of three. We have measured $\eff$ to be a factor of $8-12$ larger than their best estimate in more than a dozen GMCs in the Milky Way. Furthermore, these same GMCs seen to be in the process of disruption, so that they are at the end of their lifetimes, suggesting that {\em every} massive GMC will experience a similar burst of rapid star formation. Since none of the massive GMCs in the catalogs we have considered have very low virial parameters (or very high Mach numbers), there is no indication that these objects have extreme values of either $\alpha_{\rm vir}$ or ${\cal M}$; we conclude that supersonic turbulence is not the dominant determinant of the star formation rate in Milky Way GMCs.

Rapid star formation rates (per free-fall time) have implications for star formation in external galaxies. \citet{2010ApJ...709..191M} modeled star formation in clump galaxies, which contain star forming complexes with $R\sim0.2-1\kpc$ and $M\sim 10^8-10^9M_\odot$ ('clumps').  \citet{2010ApJ...709..191M} found that the clumps should be disrupted by radiation pressure, with $\eg\sim0.3$. Since the free fall time of the clump is several times longer than $\tms$, the implication was that $\eff\approx0.3$ or larger. \citet{2010MNRAS.tmp..635K} state that this is ``an extraordinarily high value of $\eff$''. We have seen that rate of star formation is achieved by a number of GMCs in the Milky Way, which also have free-fall times several times longer than $\tms$. Since both the Milky Way and the clump forming galaxies appear to have marginally stable disks, and to have star formation rates consistent with the Kennicutt value (although in the latter case the evidence is rather slim), high values of $\eff$ might be expected in external galaxies. A concomitant result is that the star formation efficiency of clump galaxies may well saturate at $\eg\sim0.3$, rather than proceeding to $\eg\approx1$, as envisioned by \citet{2010MNRAS.tmp..635K}.

\section{CONCLUSIONS}
We have presented four major observational results in this paper.
First, we have estimated a lower limit to the fraction of gas in a massive Milky Way GMC that will be converted into stars over the lifetime of that GMC, finding $\eg\approx0.08$. Second, we have measured the star formation rate per free-fall time, $\eff\approx0.16$, for our ionizing luminosity-selected sample of GMCs. Third, we have estimated the lifetime of massive ($\Mg\approx10^6M_\odot$ Milky Way GMCs, finding that they live $15-20\Myr$, or one to two free-fall times. Finally, we have shown that many if not most of these clouds are in the process of being disrupted by the radiation pressure of the star clusters they have given birth to.

The strong implication of these results is that the rate of star formation in Milky Way GMCs, and by extension, in the Galaxy as a whole, is not set by the properties of supersonic turbulence. 

%--------------------------------------------------------------------

\acknowledgments  
This research has made use of NASA's Astrophysics Data
System. The author is supported in part by the Canada Research Chair
program and by NSERC of Canada.

\appendix
\section{DISTRIBUTION FUNCTIONS \label{appendix:distribution functions}}
The GMC mass distribution function is generally fit by a simple power
law
\be  %$
{dN\over dm} \sim m^{-\alpha},
\ee  %$
with $\alpha\approx1.5$
\citep{1984A&A...133...99C,1986ApJ...305..892D,
1996ApJ...458..561D,2009ApJS..182..131R}. 
These papers show that Milky Way GMCs have masses $\Ml<m<\Mu$ with $\Ml\approx10^3M_\odot$ and $\Mu\approx6\times10^6M_\odot$.  The total mass in molecular gas is $M_g=10^9M_\odot$ \citep{1993AIPC..278..267D}; we assume that it is all in GMCs. 

It is useful to employ the natural logarithm of the mass as the independent variable
\be  %$
{dN\over d\ln m}=N_U\left({\Mu\over m}\right)^\beta,
\ee  %$
where $\beta=\alpha-1$. The quantity $N_U$ is approximately the number
of clouds in the mass range from $\Mu/2$ to $\Mu$. 
Using this notation, the total mass in GMCs is related to the upper and lower mass cutoff, and the power law index, by
\bea  %$
M_{Tot}&=&\int_{\Ml}^{\Mu} m{dN\over d\ln m} {d\ln m}\\
&=&{N_UM_U\over 1-\beta}
\left[1-\left({\Ml\over \Mu}\right)^{1-\beta}\right].
\eea  %$
The observed quantities are $M_{Tot}$, $\Mu$, $\Ml$, and the power law
slope $\alpha$, so we solve for $N_U$ in terms of these three:
\be  \label{eqn:Nu}%$
N_U={(2-\alpha)\left({M_{Tot}\over M_U}\right)
\over 1-\left({\Ml\over \Mu}\right)^{2-\alpha}}
\approx\gamma\left({M_{Tot}\over M_U}\right),
\ee  %$
where $\gamma\equiv (2-\alpha)$.
Using the numerical values given above, we find
\be  %$
N_U\approx 100.
\ee  %$

The  number of GMCs with mass larger than $m$ is
\be  %$
N(>m)=\int_{m}^{\Mu} {dN\over \ln m} {d\ln m}.
\ee  %$
Performing the integral we find
\be  %$
N(>m)={N_U\over \alpha-1}
\left[\left({\Mu\over m}\right)^{\alpha-1}-1\right].
\ee  %$
From equation (\ref{eqn:Nu}),
\be  %$
N(>m)=\left({2-\alpha\over \alpha-1}\right)
{\left[\left(\Mu\over m\right)^{\alpha-1}-1\right]
\over
\left[1-\left({\Ml\over\Mu}\right)^{2-\alpha}\right]}
\left({M_{Tot}\over \Mu}\right).
\ee  %$

The total number of clouds is $N_{Tot}\approx13,000$.

Next we calculate the star formation rate of the Galaxy under the assumption that 
\be  %$
\dot m_*=\eff {\Mg\over\tff}
\ee  %$
with a fixed $\eff$, i.e., the same value of $\eff$ is used for all GMCs. The total star formation rate is
\be  %$
\dot M_*=\int_{\Ml} ^{\Mu} \dot m_* {dN\over \ln dm} d\ln m.
\ee  %$
To evaluate $\tff$ we will use the fact that the surface density $\Sigma_0\equiv\Mg/(\pi\Rg^2)$ of GMCs is constant, independent of $\Mg$, e.g.,  \citet{1987ApJ...319..730S}. Then $\tff(\Mg)=\tff(\Mu)(\Mg/\Mu)^{1/4}$ and
\be  %$
\dot M_*={1\over \left({7/4}-\alpha\right)}{\eff N_U\Mu\over \tff(\Mu)}
\left[1-\left({\Ml\over\Mu}\right)^{\left({7/4}-\alpha\right)}\right]
\approx 2 {\eff M_{Tot}\over\tff(\Mu)}.
\ee  %$

Letting $\delta=(7/4) - \alpha$, the fraction of star formation that takes place in GMCs with masses greater than $m$ is
\be  %$
f_*(>m)={1-\left(m/\Mu\right)^\delta
\over
1-\left(\Ml/\Mu\right)^\delta
}
\ee  %$

It follows that the GMC mass $M_{f_*}$ at which the cumulative star formation rate reaches a fraction $f_*$ of the total star formation rate is
\be  %$
M_{f_*}=\left[1-f_*\left(1-\left({\Ml\over\Mu}\right)^\delta\right)\right]^{1/\delta} \Mu.
\ee  %$
%

%------------------------------------------------------------------------------------------------

\clearpage
\begin{deluxetable}{lccccccccccccc} 
%\rotate 
%\tablewidth{0pt} \rotate \tablecaption{STARFORMING COMPLEXES AND GIANT MOLECULAR CLOUD IDENTIFICATIONS}
\tablewidth{0pt} \tablecaption{STARFORMING COMPLEXES AND GIANT MOLECULAR CLOUD IDENTIFICATIONS}
\tablehead{ 
\colhead{catalog} & \colhead{$l$} & \colhead{$b$} & \colhead{$v_r$} & \colhead{$D$} & \colhead{$R_{\rm SFR}$} & \colhead{$f_\nu$} & 
\colhead{ref} & \colhead{catalog} & \colhead{$l$} & \colhead{$b$} & \colhead{$v_r$} & \colhead{$R_{\rm GMC}$} & \colhead{$M_{\rm GMC}$}     \\ 
\colhead{number} & \colhead{deg} & \colhead{deg} & \colhead{$\kms$} &  \colhead{$\kpc$} &  \colhead{$\pc$} & \colhead{Jy} & 
\colhead{\phantom} & \colhead{number} & \colhead{deg} & \colhead{deg} & \colhead{$\kms$} & \colhead{pc} & \colhead{$M_\odot$} \\ 
} 
 \startdata
 1 &     10.156 &  -0.384 &      15 &    14.5 &    25.0 &    29 &   1 &   14 &    10.00 &   -0.04 &      32 &    39.0 & 1.3e+05  \\ 
2 &     10.288 &  -0.136 &      10 &    15.2 &    25.0 &   150 &   1 &   15 &    10.20 &   -0.30 &       8 &    29.0 & 7.5e+05  \\ 
3 &     10.450 &   0.021 &      69 &     6.1 &     5.0 &    50 & \nodata & \nodata & \nodata & \nodata & \nodata & \nodata & \nodata  \\ 
4 &     10.763 &  -0.498 &      -1 &    16.7 &    46.0 &    29 &   1 &   16 &    10.60 &   -0.40 &      -2 &    76.0 & 2.8e+05  \\ 
5 &     22.991 &  -0.345 &      76 &    10.8 &    49.0 &   124 &   2 &   97 &    23.00 &   -0.36 &      77 &    83.4 & 2.1e+06  \\ 
6 &     23.443 &  -0.237 &     104 &     9.2 &     5.0 &   482 &   2 &  100 &    23.39 &   -0.23 &     100 &    23.3 & 4.1e+05  \\ 
7 &     23.846 &   0.152 &      95 &     5.8 &    12.0 &   138 &   2 &  105 &    23.96 &    0.14 &      81 &     8.2 & 2.1e+04  \\ 
8 &     24.050 &  -0.321 &      85 &    10.3 &    61.0 &    55 &   2 &  106 &    24.21 &   -0.04 &      88 &    22.7 & 8.2e+04  \\ 
9 &     24.133 &   0.438 &      98 &     9.7 &     4.0 &    41 &   2 &  116 &    24.63 &   -0.14 &      84 &    25.8 & 6.9e+04  \\ 
10 &     24.911 &   0.134 &     100 &     6.1 &    43.0 &   179 &   2 &  118 &    25.18 &    0.16 &     103 &    23.3 & 1.2e+04  \\ 
11 &     25.329 &  -0.275 &      63 &     4.1 &     7.0 &   248 &   2 &  121 &    25.40 &   -0.24 &      58 &    52.0 & 5.9e+05  \\ 
12 &     25.992 &   0.119 &     106 &     8.8 &    28.0 &   110 &   2 &  125 &    25.72 &    0.24 &     111 &    42.7 & 1.5e+05  \\ 
13 &     28.827 &  -0.230 &      88 &     5.3 &     2.0 &   119 &   2 &  154 &    28.80 &   -0.26 &      88 &    16.3 & 3.2e+04  \\ 
14 &     29.926 &  -0.049 &      97 &     8.7 &    13.0 &   594 &   2 &  162 &    29.89 &   -0.06 &      99 &    34.9 & 8.3e+05  \\ 
15 &     30.456 &   0.443 &      58 &     3.6 &     3.0 &    36 &   2 &  165 &    30.41 &    0.46 &      45 &     5.0 & 1.7e+03  \\ 
16 &     30.540 &   0.022 &      44 &    11.9 &     3.0 &    65 &   2 &  168 &    30.57 &   -0.02 &      41 &    44.7 & 2.7e+05  \\ 
17 &     30.590 &  -0.024 &      99 &     6.3 &    35.0 &   753 &   2 &  171 &    30.77 &   -0.01 &      94 &    41.8 & 1.1e+06  \\ 
18 &     32.162 &   0.038 &      96 &     8.2 &    12.0 &    29 &   2 &  182 &    32.02 &    0.06 &      97 &    37.7 & 1.6e+05  \\ 
19 &     34.243 &   0.146 &      37 &     2.2 &     5.0 &   130 & \nodata & \nodata & \nodata & \nodata & \nodata & \nodata & \nodata  \\ 
20 &     35.038 &  -0.490 &      51 &     3.0 &     3.0 &   130 &   1 &  200 &    34.70 &   -0.70 &      46 &    35.0 & 1.6e+06  \\ 
21 &     35.289 &  -0.073 &      48 &    11.0 &    62.0 &    25 & \nodata & \nodata & \nodata & \nodata & \nodata & \nodata & \nodata  \\ 
22 &     37.481 &  -0.384 &      50 &    10.5 &    74.0 &   130 &   2 &  211 &    37.76 &   -0.21 &      63 &    20.8 & 7.7e+04  \\ 
23 &     38.292 &  -0.021 &      57 &     3.5 &    10.0 &   114 & \nodata & \nodata & \nodata & \nodata & \nodata & \nodata & \nodata  \\ 
24 &     49.083 &  -0.306 &      68 &     5.6 &    11.0 &   229 &   1 &  233 &    49.50 &   -0.40 &      57 &    52.1 & 3.3e+06  \\ 
25 &     49.483 &  -0.343 &      60 &     5.7 &    18.0 &   229 &   2 &  234 &    49.75 &   -0.55 &      68 &    13.0 & 9.0e+04  \\ 
26 &    283.883 &  -0.609 &       0 &     4.0 &    63.0 &   848 &   3 &    7 &   283.80 &    0.00 &      -5 &    65.0 & 1.3e+06  \\ 
27 &    290.873 &  -0.742 &     -22 &     3.0 &    27.0 &   456 & \nodata & \nodata & \nodata & \nodata & \nodata & \nodata & \nodata  \\ 
28 &    291.563 &  -0.569 &      16 &     7.4 &    67.0 &   232 &   3 &   17 &   291.60 &   -0.40 &      15 &    73.0 & 1.4e+06  \\ 
29 &    298.505 &  -0.522 &      24 &     9.7 &    69.0 &   313 &   3 &   26 &   298.80 &   -0.20 &      25 &   157.0 & 8.4e+06  \\ 
30 &    310.985 &   0.409 &     -51 &     3.5 &     3.0 &    56 & \nodata & \nodata & \nodata & \nodata & \nodata & \nodata & \nodata  \\ 
31 &    311.513 &  -0.027 &     -55 &     7.4 &    56.0 &   590 &   4 &    7 &   311.70 &    0.00 &     -50 &   118.0 & 1.2e+06  \\ 
32 &    311.650 &  -0.528 &      34 &    13.6 &    29.0 &   114 &   3 &   35 &   311.30 &   -0.30 &      27 &   210.0 & 1.3e+07  \\ 
33 &    327.436 &  -0.058 &     -60 &     3.7 &    36.0 &   977 &   4 &   25 &   327.00 &   -0.40 &     -60 &    60.0 & 3.4e+05  \\ 
34 &    332.809 &  -0.132 &     -52 &     3.4 &    39.0 &  1192 &   4 &   33 &   332.60 &    0.20 &     -47 &    56.0 & 6.6e+05  \\ 
35 &    333.158 &  -0.076 &     -91 &     5.5 &     4.0 &   596 &   4 &   35 &   333.90 &   -0.10 &     -89 &   111.0 & 1.2e+06  \\ 
36 &    336.484 &  -0.219 &     -88 &     5.4 &     6.0 &   356 & \nodata & \nodata & \nodata & \nodata & \nodata & \nodata & \nodata  \\ 
37 &    336.971 &  -0.019 &     -74 &    10.9 &    49.0 &   365 &   4 &   38 &   337.10 &   -0.10 &     -74 &   162.0 & 4.4e+06  \\ 
38 &    337.848 &  -0.205 &     -48 &     3.5 &    18.0 &   750 &   4 &   39 &   337.80 &    0.00 &     -56 &   141.0 & 4.0e+06  \\ 
39 &    338.412 &   0.120 &     -33 &     2.5 &     4.0 &   640 & \nodata & \nodata & \nodata & \nodata & \nodata & \nodata & \nodata  \\ 
40 &    338.888 &   0.618 &     -63 &     4.4 &    11.0 &   128 &   4 &   41 &   339.00 &    0.70 &     -60 &    38.0 & 2.4e+05  \\ 
\enddata 
%\tablenotetext{a}{Any notes go here} 
 \tablecomments{Properties of Milky Way star forming complexes and their host GMCs. Column one gives the Rahman \& Murray (2010) catalog number of the star forming complex. The next six columns give star formation complex properties: the Galactic longitude $l$ (col. 2), latitude $b$ (col. 3), heliocentric radial velocity $v_r$ (col. 4), heliocentric kinematic distance $D$ (col. 5), the radius of the complex (col. 6) and free-free flux (col. 7). The eighth column gives the reference to the relevant GMC catalog. Column nine gives the catalog number of the GMC, followed by the GMC properties in the next five columns: $l$ (col. 10) $b$ (col. 11), radial velocity (col. 12), GMC radius (col. 13) and mass (col. 14). References: (1) \citet{1987ApJ...319..730S} (2) \citet{2009ApJ...699.1092H} (3) \citet{1987ApJ...315..122G} (4) \citet{1989LNP...331..139B}. The masses and radii listed are those in the original publications; in the calculations described in this paper, and in table 2, the GMC radii have been adjusted to $R_0=8.5\kpc$ and the mass have been adjusted as described in the text.} 
\end{deluxetable} 

\clearpage
\clearpage
\begin{deluxetable}{lcccccccccc}
\tablewidth{0pt} \tablecaption{STARFORMING COMPLEX EFFICIENCIES}
\tablehead{
\colhead{catalog} & \colhead{$L_\nu$} & \colhead{$Q$} & \colhead{$M_*$} & \colhead{$\Mg$} & \colhead{$\tff$} & \colhead{$\eg$} & \colhead{$\eff$} & \colhead{$F_{rad}/F_{Grav} $} \\ 
\colhead{number} & \colhead{$\ergs\Hz^{-1}$} &  \colhead{$\s^{-1}$ (dust adjusted)} & \colhead{$M_\odot$} & \colhead{$M_\odot$} & \colhead{$\Myr$} & \colhead{\phantom} & \colhead{\phantom} & \colhead{\phantom} \\ 
}
\startdata
1 &  7.27e+24 &  1.32e+51 &  2.10e+04 &  4.77e+05 &     24.71 &     0.042 &     0.267 &      5.79  \\ 
2 &  4.13e+25 &  7.53e+51 &  1.19e+05 &  2.09e+06 &      5.00 &     0.054 &     0.069 &      0.55  \\ 
4 &  9.64e+24 &  1.76e+51 &  2.79e+04 &  8.56e+05 &     38.14 &     0.032 &     0.308 &      6.28  \\ 
5 &  1.72e+25 &  3.14e+51 &  4.99e+04 &  4.12e+06 &      6.11 &     0.012 &     0.019 &      0.12  \\ 
6 &  4.86e+25 &  8.86e+51 &  1.41e+05 &  8.11e+05 &      2.03 &     0.148 &     0.077 &      0.68  \\ 
7 &  5.53e+24 &  1.01e+51 &  1.60e+04 &  4.87e+04 &      2.19 &     0.247 &     0.139 &      3.66  \\ 
8 &  6.96e+24 &  1.27e+51 &  2.01e+04 &  1.67e+05 &      4.50 &     0.107 &     0.124 &      2.32  \\ 
9 &  4.60e+24 &  8.38e+50 &  1.33e+04 &  1.30e+05 &      5.49 &     0.093 &     0.131 &      2.79  \\ 
10 &  7.94e+24 &  1.45e+51 &  2.30e+04 &  1.61e+04 &      8.15 &     0.587 &     1.228 &    131.91  \\ 
11 &  4.97e+24 &  9.06e+50 &  1.44e+04 &  4.35e+05 &      2.08 &     0.032 &     0.017 &      0.17  \\ 
12 &  1.02e+25 &  1.85e+51 &  2.94e+04 &  3.11e+05 &      8.70 &     0.086 &     0.193 &      3.57  \\ 
13 &  3.98e+24 &  7.26e+50 &  1.15e+04 &  6.30e+04 &      4.33 &     0.155 &     0.172 &      4.62  \\ 
14 &  5.36e+25 &  9.77e+51 &  1.55e+05 &  1.69e+06 &      2.71 &     0.084 &     0.058 &      0.41  \\ 
15 &  5.56e+23 &  1.01e+50 &  1.61e+03 &  4.29e+03 &      4.10 &     0.273 &     0.287 &     21.61  \\ 
16 &  1.10e+25 &  2.00e+51 &  3.17e+04 &  5.38e+05 &      6.72 &     0.056 &     0.096 &      1.31  \\ 
17 &  3.56e+25 &  6.49e+51 &  1.03e+05 &  2.43e+06 &      3.32 &     0.041 &     0.035 &      0.22  \\ 
18 &  2.32e+24 &  4.24e+50 &  6.72e+03 &  3.22e+05 &      6.99 &     0.020 &     0.037 &      0.58  \\ 
20 &  1.39e+24 &  2.54e+50 &  4.03e+03 &  1.73e+06 &      1.76 &     0.002 &     0.001 &      0.01  \\ 
22 &  1.71e+25 &  3.11e+51 &  4.94e+04 &  1.65e+05 &      4.27 &     0.230 &     0.252 &      5.39  \\ 
24 &  8.56e+24 &  1.56e+51 &  2.48e+04 &  3.33e+06 &      2.08 &     0.007 &     0.004 &      0.02  \\ 
25 &  8.87e+24 &  1.62e+51 &  2.57e+04 &  1.55e+05 &      1.57 &     0.142 &     0.057 &      0.80  \\ 
26 &  1.62e+25 &  2.95e+51 &  4.68e+04 &  1.12e+06 &      6.41 &     0.040 &     0.066 &      0.68  \\ 
28 &  1.51e+25 &  2.76e+51 &  4.38e+04 &  1.20e+06 &      7.35 &     0.035 &     0.066 &      0.70  \\ 
29 &  3.51e+25 &  6.40e+51 &  1.02e+05 &  7.22e+06 &      9.47 &     0.014 &     0.034 &      0.21  \\ 
31 &  3.85e+25 &  7.02e+51 &  1.11e+05 &  1.71e+06 &     37.80 &     0.061 &     0.592 &      9.84  \\ 
32 &  2.51e+25 &  4.58e+51 &  7.27e+04 &  1.12e+07 &     11.77 &     0.006 &     0.019 &      0.11  \\ 
33 &  1.59e+25 &  2.91e+51 &  4.61e+04 &  2.61e+05 &     13.87 &     0.150 &     0.533 &     13.12  \\ 
34 &  1.64e+25 &  2.99e+51 &  4.75e+04 &  5.51e+05 &      9.74 &     0.079 &     0.198 &      3.12  \\ 
35 &  2.15e+25 &  3.92e+51 &  6.22e+04 &  1.03e+06 &     20.70 &     0.057 &     0.303 &      4.86  \\ 
37 &  5.17e+25 &  9.42e+51 &  1.50e+05 &  3.53e+06 &     17.86 &     0.041 &     0.186 &      1.85  \\ 
38 &  1.10e+25 &  2.00e+51 &  3.17e+04 &  9.61e+05 &      4.55 &     0.032 &     0.037 &      0.36  \\ 
40 &  2.95e+24 &  5.38e+50 &  8.54e+03 &  2.04e+05 &      9.18 &     0.040 &     0.095 &      1.96  \\ 
\enddata
% Q_total = 1.1e+53  Q_total in known GMCs =1.0e+53
%\tablenotetext{a}{Any notes go here}
\tablecomments{Efficiencies of Milky Way star forming complexes and their host GMCs. Column one gives the Rahman \& Murray (2010) catalog number of the star forming complex. The rest of the columns give star formation complex properties; free-free luminosity $L_\nu$ (col. 2), ionizing photons emitted per second $Q$ (col. 3), stellar mass (col. 4),  GMC mass adjusted from the original as described in the text (col. 5), $\tau_{ff}$ (col. 6), GMC efficiency (col. 7), $\eff$ (col. 8), and radiation force divided by the force of gravity (col. 9).
} 
\end{deluxetable}

\clearpage

\bibliography{GMC_Life}{}

\end{document}